\documentclass[12pt]{article}
\usepackage{amsmath, amssymb}
\usepackage{graphicx}
\begin{document}
\def\bfV{{\boldsymbol V}\!}
\def\beq{\begin{eqnarray}}
\def\eeq{\end{eqnarray}}
\newcommand{\gsim}{ \mathop{}_{\textstyle \sim}^{\textstyle >} }
\newcommand{\lsim}{ \mathop{}_{\textstyle \sim}^{\textstyle <} }
\newcommand{\vev}[1]{ \left\langle{#1} \right\rangle}
\newcommand{\bra}[1]{ \langle{#1} | }
\newcommand{\ket}[1]{ | {#1} \rangle}
\newcommand{\EV}{ {\rm eV} }
\newcommand{\KEV}{ {\rm keV} }
\newcommand{\MEV}{ {\rm MeV} }
\newcommand{\GEV}{ {\rm GeV} }
\newcommand{\TEV}{ {\rm TeV} }
\def\diag{\mathop{\rm diag}\nolimits}
\def\Spin{\mathop{\rm Spin}}
\def\SO{\mathop{\rm SO}}
\def\O{\mathop{\rm O}}
\def\SU{\mathop{\rm SU}}
\def\U{\mathop{\rm U}}
\def\Sp{\mathop{\rm Sp}}
\def\SL{\mathop{\rm SL}}
\def\tr{\mathop{\rm tr}}

\def\mbf{\boldsymbol}
\def\Kahler{K\"ahler }
\def\half{\hbox{\large ${1\over2}$}}
\def\myfrac#1#2{\hbox{\large ${#1\over#2}$}}
\def\nn{\nonumber\\}
\def\tr{{\rm tr}\,}
\def\Ln{{\rm Ln}\,}
\def\VEV#1{\left\langle#1 \right\rangle}
\def\tilde{\widetilde}
\def\mylimit#1{\ \ 
        \mathrel{\mathop{\kern0pt \llongrightarrow }\limits_{#1}}\ \ }
\def\llongrightarrow{\relbar\kern-3pt\longrightarrow}
\def\K{K\"ahler }
\def\R{{\rm R}}
\def\L{{\rm L}}
\def\calG{{\cal G}}
\def\calH{{\cal H}}
\def\calA{{\cal A}}
\def\calF{{\cal F}}
\def\calO{{\cal O}}
\def\calR{{\cal R}}
\def\Gc{G^C}
\def\calGc{{\cal G}^C}
\def\Hc{H^C}
\def\Hh{\hat H}
\def\hh{\hat h}
\def\calHh{\hat{\cal H}}
\def\GH{\Gc/\Hh}
\def\dim{{\rm dim}}
\def\dimc{{\rm dim}_{\mbf C}}
\def\NG{NG\ }
\def\bfe{{\mbf e}}
\def\bfv{{\mbf v}}
\def\bfy{{\mbf y}}
\def\bfY{{\mbf Y}}
\def\sdet{{\rm det}}
\def\short#1{\kern.03em{#1}\kern.08em}
\def\slash#1{\ooalign{\hfil/\hfil\crcr$#1$}}
\def\itslash#1{\ooalign{\hfil{\it/}\hfil\crcr$#1$}}
\def\sqr#1#2{{\vcenter{\hrule height.#2pt
      \hbox{\vrule width.#2pt height#1pt \kern#1pt
          \vrule width.#2pt}
      \hrule height.#2pt}}}
\def\ssquare{{\mathchoice{\sqr34}{\sqr34}{\sqr{2.1}3}{\sqr{1.5}3}}}
\def\square{{\mathchoice{\sqr84}{\sqr84}{\sqr{5.0}3}{\sqr{3.5}3}}\,}
\def\abs#1{\left|{#1}\right|}


\baselineskip 0.7cm

\begin{titlepage}

\begin{flushright}
IPMU 10-0052 \\
YITP-10-24
\end{flushright}

\vskip 1.35cm
\begin{center}
{\large \bf
Coupling Supersymmetric Nonlinear Sigma Models to Supergravity
}
\vskip 1.2cm
Taichiro Kugo${}^{1,2}$ and  Tsutomu T.~Yanagida${}^{2,3}$
\\
\vskip 8mm
{\small 
 ${}^1$ Yukawa Institute for Theoretical Physics, Kyoto University, Kyoto 606-8502, Japan 
\\ \vskip2mm
 ${}^2$ Institute for the Physics and Mathematics of the Universe (IPMU),
University of Tokyo, Chiba 277-8568, Japan
\\ \vskip2mm
 ${}^3$ Department of Physics, University of Tokyo,
Tokyo 113-0033, Japan 
}

\vskip 1.5cm

\abstract{It is known that supersymmetric nonlinear sigma models for the
compact \Kahler manifolds $G/H$ cannot be consistently coupled to 
supergravity, since the \Kahler potentials are not invariant under the 
$G$ transformation. We show that the supersymmetric 
nonlinear sigma models can be deformed such that the \Kahler potential 
be exactly $G$-invariant if and only if one enlarges the manifolds 
by dropping all the $U(1)$'s in the unbroken subgroup $H$. Then, 
those nonlinear sigma models can be coupled to supergravity without 
losing the $G$ invariance.
}

\vskip 3cm
(Submitted to {\sl Progress of Theoretical Physics})
\end{center}
\end{titlepage}

\setcounter{page}{2}

\section{Introduction}

One of the fundamental questions in particle physics is why nature 
chooses three families of quarks and leptons but not more than that. 
Supersymmetric (SUSY) nonlinear sigma models may provide an answer to 
this question, since fermion partners of Nambu-Goldstone (NG) bosons may 
be identified \cite{BLPY} with all quarks and leptons in the standard 
model. The number of families is determined by the geometry of a given 
coset-space $G/H$. In fact, a SUSY $E_7/SU(5)\times U(1)^3$ nonlinear sigma 
model is known to accommodate three families of quarks and leptons 
\cite{KY}. It will be remarkable that if one goes to larger nonlinear 
sigma models using $E_8$, one obtains a pair of extra family and 
anti-family in addition to the three families and hence the net number 
of families remains three.

For general SUSY nonlinear sigma models, manifolds consisting of the NG 
bosons should be complex manifolds. If $G/H$ itself is a \Kahler 
manifold, the \Kahler potential for the NG chiral multiplets is uniquely 
determined by the geometry of the $G/H$. 
The \Kahler 
potential $K$ transforms as $K(\phi,\phi^\dagger) \rightarrow 
K(\phi,\phi^\dagger)+ F(\phi) + F^\dagger(\phi^\dagger)$ under the global 
$G$ transformation, where the $F$ is holomorphic function of the NG 
superfields $\phi_i$. Since the Lagrangian is given by 
$\int d^2\theta d^2\bar\theta  K(\phi,\phi^\dagger)$ in the 
rigid SUSY theory, it is invariant under the global $G$ 
transformation.

However, the supergravity (SUGRA) lagrangian is given by \cite{CFGVP,SUGRA}
\begin{equation}
\Bigl[\,\Sigma\Sigma ^\dagger\, e^{-K(\phi,\phi^\dagger)}\, \Bigr]_D 
+ \Bigl[\,\Sigma^3\, W(\phi)\,\Bigr]_F
\end{equation}
where $\Sigma$ is the chiral compensator of Weyl weight one 
in the superconformal tensor 
calculus which leads to the so-called old minimal Poincar\'e 
supergravity, and $[\, V^{(w=2)}\,]_D$ and 
$[\, \Phi^{(w=3)}\,]_F$ are the $D$-term and $F$-term 
superconformal invariant action formulae applicable to 
general-type vector multiplet $V^{(w=2)}$ of Weyl weight two and 
chiral multiplet $\Phi^{(w=3)}$ of Weyl weight three, 
respectively \cite{rf:KU}.  
The $G$ invariance is maintained by a $\Sigma$ rotation $\Sigma\rightarrow\Sigma e^{F}$ 
provided that the superpotential $W$ vanishes \cite{BW}. 
This is the case for pure nonlinear sigma model system. 
Actually, however, we also have matter fields 
besides the nonlinear sigma-model fields which usually appear in the 
superpotential term. In the real world, moreover, 
we need at least a constant term in $W$ to get a (almost) vanishing 
cosmological constant. Then, the $G$ invariance is explicitly broken 
and the NG bosons get the masses of the order of gravitino mass 
$m_{3/2}$. This is the problem whenever non-vanishing 
$F(\phi)$ appears in the $G$-transformation of the \K potential 
$K(\phi,\phi^\dagger)$. This is, in particular, the case when 
$G/H$ is a compact K\"ahler manifold. 

Nibbelink and van Holten \cite{GrootNibbelink:1998tz} 
have proposed a method for making the 
\K potential $G$-invariant to circumvent the above problem. They have added 
matter fields $S_i$ and assigned 
a $U(1)$ charge $q_i$ to them such that the superpotential $W(S)$ carries 
charge three. Then, after K\"ahler-Weyl transformation 
$\Sigma^3W(S)\rightarrow\Sigma^3$, 
the \Kahler potential becomes 
\begin{equation}
K_{\rm NL}(\phi,\phi^\dagger) + {1\over3}\ln \abs{W(S)}^2 
\end{equation}
which become $G$-invariant if the matter fields are assumed to 
receive an additional transformation under $G$-transformation:
\begin{equation}
\delta_G^{\rm additional} S_i = -q_iF(\phi) S_i \quad \Rightarrow\quad 
\delta_G W(S) = -3 F(\phi) W(S)
\end{equation} 
where $F(\phi)$ is the holomorphic shift of 
the nonlinear sigma model \K potential under $G$-transformation: 
$\delta_G K_{\rm NL}(\phi,\phi^\dagger)=F(\phi)+ F^\dagger(\phi^\dagger)$. 
The matter kinetic terms, $S^\dagger_i e^{q_iK_{\rm NL}(\phi,\phi^\dagger)}S_i$, or any 
function of them, should also be added in the total \K potential 
$K(\phi,\phi^\dagger,S, S^\dagger)$.    
If the superpotential $W(S)$ here 
is replaced by a single $H$-singlet matter $S$, then this proposal 
becomes equivalent to the recent proposal by Komargodski and Seiberg \cite{KS}. 

The problem of this proposal is, however, that it turns out 
\cite{rf:IKY} that 
the supersymmetry has to be spontaneously broken in order for 
the linear term in the nonlinear sigma model \K potential 
$K_{\rm NL}(\phi,\phi^\dagger )$ to exist nonvanishingly 
at the stationary point of the scalar potential. 

In this paper, therefore, we propose another way to solve the problem, 
which works irrespectively of the spontaneous SUSY breaking. 
We show that if one drops all the $U(1)$'s in the unbroken subgroup 
$H$ of the \Kahler manifold 
$G/H$, one can construct a SUSY nonlinear sigma model with an exactly 
$G$-invariant \Kahler potential which can be safely coupled to SUGRA.
We also show that the non-invariance under 
$G$-transformation has the same origin as the inconsistency problem of 
the Fayet-Iliopoulos terms in SUGRA. 


\section{Invariant \Kahler potential}

BKMU have presented three prescriptions, A-type to C-type, for constructing 
the invariant action for the supersymmetric system of 
nonlinear realization of $G/H$ \cite{rf:BKMUI}. 
They have also shown that there exist no A-type and C-type invariant 
actions and 
only the B-type action is available for the K\"ahlerian $G/H$ case. 
The B-type \Kahler potential is a linear combination of the terms taking 
the form
\begin{equation}
K(\phi,\phi^\dagger )=\ln\sdet_{\eta}(\xi^\dagger\xi)
\end{equation}
which is generally non-invariant under non-linearly realized $G/H$ 
transformation, but yields the so-called \K transformation 
$K(\phi,\phi^\dagger )\rightarrow K(\phi,\phi^\dagger )+F(\phi)+F^\dagger(\phi^\dagger )$. 
We examine which conditions are necessary and sufficient for the K\"ahler $G/H$ 
in order for the holomorphic function $F(\phi)$ to vanish always 
for $G$-transformations. 
We discuss, in this paper, only the B-type action formulas, since 
phenomenologically interesting nonlinear sigma models are mostly the cases 
of K\"ahler manifold $G/H$, like $E_7/SU(5)\times U(1)^3$ as 
discussed in the introduction. Moreover this is also sufficient 
since the other A-type and C-type actions, if any,  
already give invariant \Kahler potentials. 

\subsection{Complex unbroken subgroup $\Hh$}

We first need to recall the BKMU construction \cite{rf:BKMUI}. When the symmetry 
$G$ of the system is spontaneously broken to 
its subgroup $H$, then the generators $T_A$ of $G$ are divided into two 
parts, $S_\alpha$ of the unbroken subgroup $H$ and $X_a$ of the rest:
\begin{equation}
\{\ T_A\in\calG\ \} = \{\  S_\alpha\in\calH, \quad X_a\in\calG\!-\!\calH \ \}.
\label{eq:G/Hdecomp}
\end{equation}
Since this is a decomposition from the real group view point, these 
generators are 
{\it hermitian}; $S_\alpha^\dagger=S_\alpha$ and $X_a^\dagger=X_a$. 
To each of the independent broken generators $X_a$, 
a NG boson (real field) $\pi^a$ corresponds. Now, in 
supersymmetric theory, this spontaneous breaking $G\rightarrow H$ 
correspond to the breaking of complex group $\Gc$ (complex extension of 
$G$, obtained by extending the real parameters of $G$ to complex values) 
into its certain complex subgroup $\Hh$ \cite{KOY}. $\Hh$ 
always contains $\Hc$, the complex extension of $H$:
\begin{equation}
\Hh \supset\Hc .
\end{equation}
Therefore, from the viewpoint of the complex group breaking $\Gc\rightarrow\Hh$, 
the real broken generators $X_a$ in Eq.~(\ref{eq:G/Hdecomp}) are 
recombined and split into complex unbroken generators $R_{\alpha'}\in\Hh$ 
and complex broken generators $\bar X_I\in\calGc\!\!-\!\calHh$:
\begin{equation}
\{\ T_A\in\calG\ \} = \{\  (S_\alpha,\ R_{\alpha'})\in\calHh, \quad 
\bar X_I\in\calGc\!\!-\!\calHh \ \}.
\label{eq:Gc/Hhdecomp}
\end{equation}

%
%
%

The basic variable of the BKMU theory is the following variable 
parameterizing the (right) coset space 
$\Gc/\Hh$: 
\begin{equation}
\xi(\phi) \equiv e^{\phi^I\bar X_I}\ , \qquad 
\hbox{$\phi^I\in{\mbf C}$: identified with NG chiral superfields}. 
\end{equation}
Action of a (real) group element 
$g\in G$ from the left of course yields an element of the 
complex group $\Gc$ since $\xi(\phi)\in\Gc$ and can be parameterized as 
\begin{eqnarray}
g\,\xi(\phi)
&=& e^{\phi'^I\bar X_I}\cdot e^{a^\alpha S_\alpha+b^{\alpha'}R_{\alpha'}} 
\qquad (a^\alpha,\ b^{\alpha'}\in{\mbf C} )\nn
&=& \xi(\phi')\cdot\hh(\phi,g) 
\qquad \qquad  \hh\in\Hh 
\end{eqnarray}
The nonlinear $G$-transformation of the NG superfields $\phi$ is then
defined by 
\begin{equation}
\xi(\phi) \mylimit{g\in G} \xi(\phi')= g\,\xi(\phi)\,\hh^{-1}(\phi,g) \ .
\label{eq:Gtransf}
\end{equation}

The problem is how to construct invariants under this $G$ transformation. 
The factor $g$ is easily cancelled by making $\xi^\dagger\xi$:
\begin{eqnarray}
\xi^\dagger(\phi)\xi(\phi) \mylimit{g\in G} && 
(\hh^{-1}(\phi,g))^\dagger\xi^\dagger(\phi)g^\dagger\cdot
g\,\xi(\phi)\,\hh^{-1}(\phi,g) \nn
 && \qquad = (\hh^{-1}(\phi,g))^\dagger\xi^\dagger(\phi)\xi(\phi)\,\hh^{-1}(\phi,g) \ .
\label{eq:Gtransfxx}
\end{eqnarray}
But, unlike in non-SUSY case, cancelling the remaining factors 
$(\hh^{-1})^\dagger$ and $\hh^{-1}$ is non-trivial since $(\hh^{-1})^\dagger$ is
the function of anti-chiral superfields $\phi^\dagger $ while $\hh^{-1}$ is
that of chiral superfield $\phi$. 
BKMU has given three recipes for doing this, but here 
we are interested only in the B-type action, which is the unique 
possibility for the case of the \K manifold $G/H$.

\subsection{B-type invariants}

For the B-type invariants, it is known to be sufficient to consider 
only the fundamental representation \cite{rf:BKMUI,rf:IKK}.   So all 
the generators and group elements are henceforth understood to be 
the representation matrices in the fundamental representation. 

Find projection matrices $\eta_i$ in the fundamental representation space 
$\bfV$ satisfying
\begin{equation}
\Hh\eta_i = \eta_i\Hh \eta_i, \qquad \eta_i^2=\eta_i, \quad \eta_i^\dagger=\eta_i.
\label{eq:projection}
\end{equation}
Then, with such projection matrices $\eta_i$ ($i=1,\,2,\,\cdots\,$) 
and arbitrary coefficients $c_i$ 
\begin{equation}
K(\phi,\phi^\dagger )= \sum_i c_i \ln\sdet_{\eta_i}(\xi^\dagger\xi)
\label{eq:KahlerPot}
\end{equation}
gives the \K\ potential 
whose $\int d^2\theta d^2\bar\theta$ integral is $G$-invariant.  
Here $\sdet_{\eta_i}$ denotes the determinant in the $\eta_i$ projected 
subspace $\eta_i\bfV$. 

We can easily show this if we use the property (\ref{eq:projection}) of 
projection matrices $\eta$: 
\begin{eqnarray}
\ln\sdet_\eta(\xi'^{\dagger}\xi') 
&=& \ln\sdet_\eta\bigl(\eta(\hh^{-1})^\dagger\xi^\dagger\xi\,\hh^{-1}\eta\bigr) \nn
&=& \ln\sdet_\eta\bigl(\eta(\hh^{-1})^\dagger\eta\cdot\eta\xi^\dagger\xi\eta\cdot
\eta\hh^{-1}\eta\bigr) \nn
&=& \ln\sdet_\eta(\hh^{-1\dagger})+\ln\sdet_\eta(\xi^\dagger\xi) +\ln\sdet_\eta(\hh^{-1}) 
\end{eqnarray} 
But the first and third terms in the last expression is functions 
of anti-chiral superfields $\phi^\dagger $ and of chiral superfields $\phi$ alone, 
respectively, so that the \K potential (\ref{eq:KahlerPot}) 
is really a quantity receiving a \Kahler transformation under the 
$G$-transformation: 
\begin{eqnarray}
&&K(\phi,\phi^\dagger )  \mylimit{g\in G} 
K(\phi',\phi^{\dagger\prime}) = K(\phi,\phi^\dagger ) + F(\phi) + F^\dagger(\phi^\dagger ), \nn
&&\hspace{1em}
\hbox{with}\quad   
F(\phi) = \sum_i c_i \ln\sdet_{\eta_i}(\hh^{-1}).
\end{eqnarray}

\subsection{$G$-invariant \K potential}

Now we can answer the question when the \K potential itself becomes 
$G$-invariant; namely, we consider when the yielded holomorphic functions 
$F(\phi)$ and their complex conjugate $F^\dagger(\phi^\dagger )$ always vanish for any 
$G$-transformation. The holomorphic function can be rewritten into the form
\begin{equation}
F(\phi)=\ln\sdet_\eta(\hh^{-1}) = -\tr\Ln(\eta\hh^{-1} \eta) 
= \tr[\,\eta\,\Ln(\hh)\,] 
= -\tr[\,\eta\,\Ln(\hh)\,] 
\label{eq:holomF}
\end{equation}
where $\Ln$ is the  logarithm function of matrix and 
the the property (\ref{eq:projection}) of 
projection matrices $\eta$ has been used in the second equality.
$\Ln(\hh)$ is an element of the Lie-algebra $\calHh$. 

The $\calHh$-structure theorem by BKMU tells us that
\begin{equation}
\calHh = \calH \oplus \calR, \qquad 
\calH \ = \ \bigoplus_i \calH_i \ \oplus \  \bigoplus_a {\cal U}(1)_a.
\end{equation} 
where $\calR$ is a nilpotent ideal of $\calHh$ whose generators are 
represented by nilpotent matrices, and $\calH$ is a direct sum of 
simple algebras $\calH_i$ and $U(1)$ algebras ${\cal U}(1)_a$
whose generators can be taken to be hermitian matrices. 
We can decompose the representation space $\bfV$ of $G$ into 
a direct sum of $\calH$-irreducible representation spaces $\bfV_i$, 
$\bfV = \bigoplus_i \bfV_i$. Since $\bfV$ is now the fundamental representation, 
the $\bfV_i$ is the fundamental representation of the simple algebra 
$\calH_i$ and all the other simple algebras $\calH_j (j\not=i)$ are trivial 
there. It may be helpful to see how the representation matrices 
for the generators look like in the base corresponding to this 
decomposition. For the case of four simple 
algebras $\calH_i\ (i=1,2,3,4)$, for instance, it looks like
\begin{equation}
\left(
\begin{array}{c|c|c|c}
\calH_1 &  \calR_{12} &  \calR_{13} &  0 \\ \hline
 0      &  \calH_2    &  \calR_{23} &  0 \\ \hline
 0      &   0         &  \calH_3    &  0 \\ \hline
 0      &   0         &   0     &  \calH_4 \\ 
\end{array}
\right)
\end{equation}
The generators of $\calH_i$ 
are represented by hermitian matrices having non-zero entries in the 
diagonal block denoted by the same letter $\calH_i$. The generators 
of the nilpotent ideal $\calR$ 
of $\calHh$ have non-zero entries at off-diagonal blocks (which, generally, 
can be placed in upper right blocks). We took an example of $\calHh$  
not containing $\calR_{i4} (i=1,2,3)$. 
The generators of $U(1)$'s have non-zero entries only at diagonal 
matrix elements, which can spread over several $\calH_i$ blocks but 
must be proportional to unit matrix in each $\calH_i$ block because of 
the irreducibility of the representation space $\bfV_i$. 

For this example, the independent projection operators $\eta_i$ 
satisfying the property 
(\ref{eq:projection}) are given by
\begin{eqnarray}
&&\eta_1=
\left(
\begin{array}{c|c|c|c}
{\bf1}_{n_1} &  0 &  0 &  0 \\ \hline
 0      &   0    &  0 &  0 \\ \hline
 0      &   0    &  0 &  0 \\ \hline
 0      &   0    &  0 &  0 \\ 
\end{array}
\right), \quad 
\eta_2=
\left(
\begin{array}{c|c|c|c}
{\bf1}_{n_1} &  0 &  0 &  0 \\ \hline
 0      &   {\bf1}_{n_2}    &  0 &  0 \\ \hline
 0      &   0    &  0 &  0 \\ \hline
 0      &   0    &  0 &  0 \\ 
\end{array}
\right), \quad 
\eta_3=
\left(
\begin{array}{c|c|c|c}
{\bf1}_{n_1} &  0 &  0 &  0 \\ \hline
 0      &   {\bf1}_{n_2}    &  0 &  0 \\ \hline
 0      &   0    &  {\bf1}_{n_3} &  0 \\ \hline
 0      &   0    &  0 &  0 \\ 
\end{array}
\right), \nn
&&\eta_4=
\left(
\begin{array}{c|c|c|c}
0 &  0 &  0 &  0 \\ \hline
 0      &   0    &  0 &  0 \\ \hline
 0      &   0    &  0 &  0 \\ \hline
 0      &   0    &  0 &  {\bf1}_{n_4} \\ 
\end{array}
\right)
\end{eqnarray}
where ${\bf1}_{n_i}$ is the $n_i$-dimensional unit matrix with 
$n_i={\rm dim}\,\bfV_i$. Note that the inclusion relation 
$\eta_1\bfV\subset\eta_2\bfV\subset\eta_3\bfV$ is determined by the structure of the 
nilpotent ideal $\calR$ of $\calHh$.

If there were no projection $\eta$ in (\ref{eq:holomF}),  then 
the holomorphic function $F(\phi)=\tr[\,\Ln(\hh)\,]$ clearly 
vanishes if $\hh$ contains no $U(1)$ parts, since the simple group 
generators and nilpotent generators are traceless. We now show 
that the same condition is required for the general \K potential 
not to yield non-zero $F(\phi)$ under any $G$ transformation.

First, the simple group generators in $\calH_i$ are always traceless 
in any representation space, so in particular, the trace restricted 
in the $\calH$-irreducible subspace $\bfV_i$ is zero. 
From the property (\ref{eq:projection}) 
of the projection operators $\eta_i$, each projected space $\eta_j\bfV$ satisfy 
$\calHh\eta_j\bfV \subset\eta_j\bfV$. So each $\calH$-irreducible subspace $\bfV_i$ is 
either fully contained in  $\eta_i\bfV$ or otherwise not contained at all. 
Therefore, the generators of the simple algebras $\calH_i$ in $\hh$  
cannot give non-zero contribution to $\tr[\,\eta_j\,\Ln(\hh)\,]$ for 
any projection operators $\eta_j$. 
It is also clear that the nilpotent generators of $\calR$ in $\hh$ 
cannot contribute either to $\tr[\,\eta_j\,\Ln(\hh)\,]$ for 
any $\eta_j$. 

So the only possibility for giving non-vanishing $\tr[\,\eta_j\,\Ln(\hh)\,]$ 
is the generators of $U(1)$-subalgebras ${\cal U}(1)_a$ contained in $\hh$.
We now show that $U(1)$-subalgebras, if contained in $\calH$, 
actually gives 
non-vanishing holomorphic contributions to $F(\phi)$ for generic 
\K potentials, so that it is necessary and sufficient for the 
generic \K potential to be exactly invariant under $G$-transformation 
that the unbroken subgroup $\Hh$ contains no $U(1)$ factor 
groups.

The generator of a $U(1)$-subalgebra ${\cal U}(1)_a$, which we denote 
as $Y_a$, are proportional to unit matrix in 
each subspace $\bfV_i$ as we noted above. 
Therefore, the trace restricted in the $\bfV_i$ subspace of any $U(1)$ 
generator $Y_a$ is non-zero; $\tr_{\bfV_i}Y_a \not=0$. But, as seen in 
the above example explicitly, and is generally true, 
the projection operator onto $\bfV_i$ is realized as one of the $\eta_j$ 
projectors, or otherwise, as the difference $\eta_i-\eta_{i-1}$, 
so that       
\begin{equation}
\tr_{\bfV_i}Y_a = \tr[ \eta_i Y_a\,] \not=0 \qquad \hbox{or}\qquad 
\tr_{\bfV_i}Y_a = \tr[ \eta_i Y_a\,]-\tr[ \eta_{i-1} Y_a\,] \not=0\,.
\end{equation}
Thus, for the generic \K potential (\ref{eq:KahlerPot}), this type of 
$\eta$-projected trace of $U(1)$ generator 
necessarily appears in the holomorphic function $F(\phi)$ in $\delta K$, and 
gives non-vanishing $F(\phi)$.  
Therefore, as far as we allow the most general form of \Kahler potential 
(\ref{eq:KahlerPot}), any $U(1)$ algebras should not be included 
in the $\Hh$. 
This finishes the proof.

A few remarks are in order: 

1) If $G/H$ is a \K manifold, then it is possible to choose 
the complex subgroup $\Hh$ such that $G/H\simeq G^C/\Hh$ and 
the boson components of the chiral superfields parameterizing $G^C/\Hh$, 
both the real and imaginary parts, all stand for the true 
NG bosons corresponding to the spontaneous breaking $G\ \rightarrow\ H$. 
This case was called {\em pure realization}. However, unfortunately, 
it is known that 
$H$ necessarily contains at least a $U(1)$ factor group for K\"ahlerian 
$G/H$ case, so that the \K potential for such a case is generally not 
invariant under $G$-transformation. 
So it cannot couple to supergravity in a $G$-invariant manner as 
far as a constant term exists in a superpotential. 

2) If we restrict the \K potential to a {\em non-generic} form, however, 
then, we can have an exact $G$-invariant \K potential even for $H$ 
containing $U(1)$ factor groups. For instance, the present authors and 
Uehara \cite{KUY} once discussed supersymmetric $U(4n+2)/SU(2)\times U(4n)$ nonlinear 
sigma model in the context of identifying weak gauge bosons as composite 
gauge fields of a hidden local symmetry. There, actually two projection 
operators exist: corresponding to the $\calH$-irredusible 
decomposition of $4n+2$-dimensional vector space into $2$-dimensional 
one and 
$4n$-dimensional one, 
\begin{equation}
\eta_1= 
\left(
\begin{array}{c|c}
{\bf1}_{2} &  0 \\ \hline
 0      &   0   \\
\end{array}
\right), \qquad 
\eta_2=
\left(
\begin{array}{c|c}
{\bf1}_{2} &  0 \\ \hline
 0      &   {\bf1}_{4n}   \\
\end{array}
\right)
\end{equation}
Actually $\eta_2$ is just the unit operator in the total vector space 
so that it is not a genuine projection operator, but it is an allowed 
projection operator giving a non-trivial \K potential in this case 
since $G=U(4n+2)=SU(4n+2)\times U(1)$ contains a $U(1)$. They 
considered the \K potential using $\eta_1$ projection operator alone, 
then the $U(1)$ factor group element contained in $H=U(4n)=SU(4n)\times U(1)$ 
does not contribute to $\tr [\eta_1\Ln(\hh) ]$ since the $U(1)$ is 
acting only in the $4n$-dimensional subspace $(1-\eta_1)\bfV$.

\section{Common origin with the Fayet-Iliopoulos term problem}

Komargodski and Seiberg \cite{KS} have already noticed the similarity 
between the difficulties in coupling the system to supergravity for 
two cases of the nonlinear sigma model and the Fayet-Iliopoulos term. 
But these two problems are not merely similar but is actually possessing
{\em the same} origin. This is because all the nonlinear $G/H$ 
sigma-model Lagrangian can be cast into the form in which the {\em 
hidden local symmetry} $H_{\rm local}$ is made manifest by introducing 
auxiliary gauge field variables. If $H$ contains $U(1)$ factor groups, 
the rewritten lagrangian with a manifest hidden-local symmetry contains 
the Fayet-Iliopoulos term for the hidden-local $U(1)$ vector superfield.

For the simplest case of Grassmannian \K manifold 
$U(n+m)/U(n)\times U(m)$, this fact has been known for a long time. 
Zumino \cite{rf:Zumino} has written the \K potential for this coset 
space in the form
\begin{equation}
K(\phi,\phi^\dagger ) = \tr \Ln ( {\bf1}_n + \phi^\dagger\phi) 
=\ln \det ( {\bf1}_n + \phi^\dagger\phi) 
\end{equation}
where $\phi$ is a chiral superfield valued $m\times n$ matrix, 
which is related to 
the BKMU coset variable $\xi\in G^C/\Hh$ by
\begin{equation}
\xi= 
\bordermatrix{     & n & m \cr
               n & {\bf1}_{n} & 0 \cr
               m &  \phi& {\bf1}_{n} \cr
            }
\end{equation} 
and Zumino's action is identical with BKMU's 
\begin{equation}
K(\phi,\phi^\dagger )= \ln\sdet_{\eta_1}(\xi^\dagger\xi)
\end{equation}
with the unique projection operator $\eta_1$ in this case:
\begin{equation}
\eta_1=
\left(
\begin{array}{c|c}
{\bf1}_{n} &  0 \\ \hline
 0      &   0  \\
\end{array}
\right)\ .
\end{equation}
Aoyama \cite{rf:Aoyama} has shown for the first time that 
this Lagrangian is equivalently rewritten into the following 
form possessing $U(n)_{\rm local}$ symmetry:
\begin{equation}
{\cal L}=\int d^2\theta d^2\bar\theta\,K(\Phi,\Phi^\dagger,V)\qquad 
K(\Phi,\Phi^\dagger,V)= \tr(\Phi^\dagger\Phi e^V) - g\,\tr V \ .
\label{eq:AoyamaLag}
\end{equation}
where $\Phi$ is an $(n+m)\times n$ matrix chiral superfield 
and $V$ is a $n\times n$ $U(n)$ gauge superfield. 
This lagrangian is manifestly invariant under 
the global $U(n+m)$ and local $U(n)$ transformation:
\begin{equation}
\Phi\ \rightarrow\ g \Phi h^\dagger(x,\theta,\bar\theta), \quad 
g\in U(n+m), \ \   
h(x,\theta,\bar\theta)
\in U(n)_{\rm local}
\end{equation}
Since the $U(n)$ gauge superfield $V$ is an auxiliary field possessing 
no kinetic term, the $V$ equation of motion can be solved:
\begin{equation}
{\delta{\cal L}\over\delta V}=0 \quad \Rightarrow\quad g\,e^{-V}= \Phi^\dagger\Phi\quad \Rightarrow\quad 
-V = \Ln(\Phi^\dagger\Phi)\ .
\end{equation}
Substituting this back into Eq.~(\ref{eq:AoyamaLag}), we find the 
\K\ potential to become
\begin{equation}
K = g\,\tr\Ln(\Phi^\dagger\Phi) = g\,\ln\det(\Phi^\dagger\Phi).
\end{equation}
This still possesses the $U(n)_{\rm local}$ gauge symmetry with 
chiral superfield parameter, so that we can take 
\begin{equation}
\Phi= \bordermatrix{
   & n        \cr
n  & {\bf1}_n \cr
m  & \phi    \cr},
\label{eq:gaugefix}
\end{equation}
as a {\it gauge fixing condition}. Then it reduces to the original 
Zumino's form of the \K potential. 

If we take this gauge condition in the Lagrangian (\ref{eq:AoyamaLag}), 
the global $G$-transformation induces a local gauge 
transformation $h(\phi,g)\in U(n)_{\rm local}$ in order to keep the 
form (\ref{eq:gaugefix}). Then it is actually the Fayet-Iliopoulos 
term $-g \tr V$ that yield the holomorphic term shift $F(\phi)=
\tr \Ln h(\phi,g)$ of the \K potential. This manifestly shows the 
equivalence of the two problems of the nonlinear sigma model and the 
Fayet-Iliopoulos term.

This type of rewriting of the Lagrangian into the form in which 
the hidden local symmetry $\calHh$ is manifest was given for more 
general \K manifold cases in Ref.\cite{rf:BKY}. So we can 
generally see the common root of the two problems.

\section{Discussion and conclusions}

The SUSY nonlinear sigma model for $E_7/SU(5)\times U(1)^3$ is very 
interesting, since it accommodates just three families of quark and 
lepton chiral multiplets as NG multiplets. However, this nonlinear sigma
model has two independent problems. One is that it suffers from 
so-called nonlinear sigma model anomalies \cite{NSM-anomaly}. That is, 
the fermion path-integral is ill defined on the \Kahler manifold. The 
other problem is that the model can not be coupled to SUGRA unless the 
superpotential vanishes, $W=0$ \cite{BW}. It was pointed out \cite{YY} 
that the former problem can be solved if one introduces one extra matter
multiplet transforming as ${\bf 5}^*$ of the unbroken $SU(5)$. In this 
paper we have shown that we can make the \Kahler potentials for 
nonlinear sigma models completely invariant under the global symmetry
$G$ if we eliminate all U(1) subgroups from the unbroken subgroup $H$. 
Then, they can be easily coupled to SUGRA without any explicit breaking 
of the $G$ symmetry. 

Now we propose a $E_7/SU(5)$ nonlinear sigma model coupled to SUGRA, 
which consists of three (${\bf 5^*+10+1+1}$) + one ${\bf 5}$ as NG 
multiplets and one extra matter ${\bf 5^*}$. We must introduce the gauge
interactions of the $SU(3)\times SU(2)\times U(1)$ subgroup of the unbroken 
$SU(5)$ and also Yukawa couplings for quark and lepton chiral multiplets
to make the model realistic. They are explicit breaking terms of the 
total group $G$. We do not know the origin of the breaking terms, but if
they are only sources of the explicit breaking we may have interesting 
predictions testable at LHC \cite{MNSY}. Namely, all squarks and 
sleptons are massless at the tree level even after the SUSY is 
spontaneously broken. This is because the $G$ invariance is kept 
unbroken in the limit of all gauge and Yukawa couplings vanishing and 
the masslessness of the NG bosons are guaranteed as long as the SUSY 
breaking sector never breaks the $G$ symmetry. 
And the introduction of 
the gauge and Yukawa interactions do not contribute to the NG boson's 
SUSY-breaking masses at the tree level. 
Therefore, it is reasonable to assume that all
squarks and sleptons are massless at the cut-off scale, say the Planck 
or GUT scale. Then, the squarks and sleptons receive their masses from 
higher order corrections from the gauge and Yukawa interactions, which 
are calculable. It is extremely interesting to test this hypothesis
at LHC.

\section*{Acknowledgments}

The authors thank K.-I.~Izawa and K.~Yonekura for useful discussions. 
T.~Kugo is grateful to the hospitality at IPMU where 
this work was performed. 
TK is supported by the Grant-in-Aid for the Global COE Program "The Next 
Generation of Physics, Spun from Universality and Emergence" from the 
Ministry of Education, Culture, Sports, Science and Technology (MEXT) of
Japan. TK is also partially supported by a Grant-in-Aid for Scientific 
Research (B) (No.\ 20340053) from the Japan Society for the Promotion of
Science. 
This work was supported by World Premier International Research
Center Initiative (WPI Initiative), MEXT, Japan.

\makeatletter
\def\@JLone<#1,#2>{#1}
\def\@JLtwo<#1,#2,#3>{#2}
\def\@JLyear<#1,#2,#3,#4>{#3}
\def\@JLpage<#1,#2,#3,#4>{#4}
\def\JL#1{\@JLone<#1>\ {\bf \@JLtwo<#1>} (\@JLyear<#1>), \@JLpage<#1>}
\def\@Jpage<#1,#2,#3>{#3}
\def\andvol#1{{\bf \@JLone<#1>} (\@JLtwo<#1>), \@Jpage<#1>}
\def\PTP#1{Prog.\ Theor.\ Phys.\ \andvol{#1}}
\def\JPSJ#1{J.~Phys.\ Soc.\ Jpn.\ \andvol{#1}}
\def\PR#1{Phys.\ Rev.\ \andvol{#1}}
\def\PRL#1{Phys.\ Rev.\ Lett.\ \andvol{#1}}
\def\PL#1{Phys.\ Lett.\ \andvol{#1}}
\def\NP#1{Nucl.\ Phys.\ \andvol{#1}}
\def\JMP#1{J.~Math.\ Phys.\ \andvol{#1}}
\def\IJMP#1{Int.\ J.~Mod.\ Phys.\ \andvol{#1}}
\def\CMP#1{Commun.\ Math.\ Phys.\ \andvol{#1}}
\def\JP#1{J.~of Phys.\ \andvol{#1}}
\def\ANN#1{Ann.\ of Phys.\ \andvol{#1}}
\def\NC#1{Nouvo Cim.\ \andvol{#1}}
\makeatother

\end{document}